\PassOptionsToPackage{pdfpagelabels=false}{hyperref}

\documentclass[usenatbib]{mnras}

\usepackage{graphicx}	
\usepackage{amsmath}	
\usepackage{amssymb}	
\usepackage{multicol}        
\usepackage{bm}		
\usepackage{pdflscape}	
\usepackage{microtype}
\usepackage{float}

\usepackage[english]{babel} 
\usepackage{placeins} 

\newcommand{\wfirst}{{\em WFIRST}}
\newcommand{\bluetides}{{\sc BlueTides}}

\usepackage[T1]{fontenc}
\usepackage{ae,aecompl}

\title[\bluetides\ Forecasts for \wfirst]{Forecasts for the  WFIRST High Latitude Survey using the {\sc BlueTides} Simulation}

\author[Dacen Waters et al.]
{Dacen Waters$^{1}$, 
Tiziana Di Matteo$^{1}$,
Yu Feng$^{2}$, 
Stephen M. Wilkins$^{3}$,
\newauthor Rupert A.C. Croft$^{1}$
\\
$^1$\,McWilliams Center for Cosmology, Physics Dept., Carnegie Mellon
University, Pittsburgh PA, 15213, USA \\
$^2$\,Berkeley Center for Cosmological Physics, University of California at
Berkeley, Berkeley, CA 94720, USA \\
$^3$\,Astronomy Centre, Department of Physics and Astronomy, University of Sussex, Brighton, BN1 9QH, UK \\
}
\date{}

\pubyear{2016}

\begin{document}
\label{firstpage}
\pagerange{\pageref{firstpage}--\pageref{lastpage}}
\maketitle

\begin{abstract}
  We use the \bluetides\ simulation to predict the properties of the
  high-$z$ galaxy and active galactic nuclei (AGN) populations for the
  planned 2200deg$^2$ Wide-Field Infrared Survey Telescope's
  (\wfirst)-AFTA High Latitude Survey (HLS). \bluetides\ is a
  cosmological hydrodynamic simulation, which incorporates a variety
  of baryon physics in a $(400h^{-1} \mathrm{Mpc})^3$ volume evolved
  to $z=8$ with 0.7 trillion particles. The galaxy luminosity
  functions in the simulation show good agreement with all the current
  observational constraints (up to $z=11$) and predicts an enhanced
  number of UV bright galaxies. At the proposed depth of the HLS ($m < 26.75$),
  \bluetides\ predicts $10^6$ galaxies at $z=8$ with a few up to
  $z\sim 15$ due to the enhanced bright end of the galaxy luminosity
  function. At $z=8$, galaxies in the mock HLS have specific star
  formation rates of $\sim 10 {\rm Gyr}^{-1}$ and ages of $\sim 80
  {\rm Myr}$ (both evolving linearly with redshift) and a non-evolving
  mass-metallicity relation. \bluetides\ also predicts $\sim 10^4$ AGN
  in \wfirst\ HLS from $z=8$ out to $z\sim 14$. These AGN host black
  holes of $M\sim 10^6-10^8 M_\odot$ accreting close to their
  Eddington luminosity. Galaxies and AGN have host halo masses of
  $M_{halo}\sim 10^{11-12} M_\odot$ and a linear bias $b\approx
  13-20$. Given the expected galaxy space densities, their high bias
  and large volume probed we speculate that it may be feasible for
  \wfirst\ HLS detect the Baryon Acoustic Oscillation peak in the
  galaxy power spectrum out to $z=8-9$.
\end{abstract}

\begin{keywords}
galaxies: high-redshift - galaxies: abundances - galaxies: evolution - galaxies: formation - dark ages, reionization, first stars
\end{keywords}
\begin{table*}
\centering
\begin{tabular}{||c c c c||} 
 \hline
 Parameter & Value & Description  \\ [0.1ex] 
 \hline\hline
$\Omega_\Lambda$ & 0.7186 & Vacuum energy density \\
$\Omega_{m}$ & 0.2814 & Matter density \\
$\Omega_b$ & 0.0464 & Baryon density\\
$h$ & 0.697 & Dimensionless Hubble parameter \\
$n_s$ & 0.971 & Spectral index \\
$\sigma_8$ & 0.82 & Linear mass dispersion at $8 h^{-1}$ Mpc \\
L & $400h^{-1}$ Mpc & Length of one edge of the simulation box\\
N & $2\times 7040^3$ & Initial number of gas and dark matter particles\\
$M_{DM}$ & $1.2\times10^7h^{-1}$ Mpc & Mass of one dark matter particle \\
$M_{Gas}$ & $2.36\times10^6h^{-1}$ Mpc & Mass of one gas particle \\
$M_{BH}$ & $5\times10^5h^{-1}$ Mpc & Seed mass of black hole particles \\
 \hline
\end{tabular}
\caption{Important parameters for the \bluetides\ simulation. \bluetides\ uses the cosmology from the Wilkinson Microwave Anisotropy Probe nine-year data release \citep{WMAP}.}
\label{table:BlueTides Parameters}
\end{table*}

\section{Introduction}
At the current high redshift observational frontier ($z\sim6-10$)
there is, excitingly, evidence for a substantial population of
galaxies (see the compilation by \citet{Bouwens} (hereafter B15) and
references therein), with glimpses of intriguing properties seen
\citep{Oesch, Holwerda}. These observations provide initial
measurements of the galaxy luminosity function (LF)
\citep[B15;][]{Oesch, McLeod} relying on around $\sim 100$ galaxies at
redshift $z=8$ but fewer beyond redshift $z=9$.  However, the 2 arcmin
wide IR field of view of HST WFC3 has meant that cosmic variance is a
dominant component of many observational studies, making clustering
measures and searches for rare objects extremely difficult. The James
Webb Space Telescope (JWST) will increase the depth to which we
observe in deep fields of view, but cosmic variance will likely still
persist. However, the Wide-Field Infrared Survey Telescope
(\wfirst-AFTA) will transform the field, resolving these issues by
providing depth comparable to that of {\em HST} Ultra Deep Fields with
a field of view comparable to that of a ground based survey.

\citet{WFIRST} (hereafter S13) reports an expected $10\sigma$ and
$5\sigma$ limiting magnitude of $26$ and $26.75$ respectively, which
is comparable to that of the {\em HST}. The \wfirst\ High Latitude
Survey (HLS) is planned to have a 2200 deg$^2$ field of view,
which will dramatically increase the number of galaxies available at
redshifts $z=8$ and above. The \wfirst\ HLS will map this large
portion of the sky in four NIR passbands (Y, J, H, and F184) and will
include a slitless spectroscopic survey component that will obtain
$R=\lambda/\Delta\lambda=600$ spectra allowing redshift measurements
for these deep field objects. The observation of these high redshift
galaxies are critical for our understanding of the first galaxies as
well as their role in the epoch of reionization.

Numerical simulations, required to make theoretical predictions for
these early times, are lacking however, particularly those with the
dynamic range to cover both the formation of individual objects and
make large scale statistical studies of them.  Over the last few
years, several large volume cosmological simulations of galaxy
formation have been performed to study structure growth in the
universe, galaxy formation, and reionization. {\it MassiveBlack} I ran
a $533 h^{-1}\rm Mpc$ side-length box to a redshift $z = 4.75$
\citep{MassiveBlack}. {\it MassiveBlack II }reduced the boxsize to
$100 h^{-1}\rm Mpc$ but had an improved resolution and ran all the way
to redshift $z=0$ \citep{Khandai}. {\it Illustris} \citep{Illustris}
and the {\it EAGLE} simulation \citep{Eagle} are both similar in size
to {\it Massive Black II} but with different subgrid physics processes
and feedback mechanisms.

With our newest simulation, \bluetides\ (\citealt{Feng}, hereafter
F16, and \citealt{Feng_2}) (and the recent radical updates to the code
efficiency, smoothed particle hydrodynamics formulation and star
formation modeling) we have reached an unprecedented combination of
volume and resolution.  This enables us to cover the evolution of most
of the galaxy mass function for the first billion years of cosmic
history. We now are able to meet the challenge of simulating the next
generation space telescope fields.

Cosmological simulations such as \bluetides\ are especially relevant
to the high redshift observational frontier. This can be seen if we
consider the recent discovery of the highest redshift galaxy to
date. \citet{Oesch16} observed a remarkably bright ($M_{\rm UV} =
-22.1$) $z=11$ galaxy, GN- $z11$, in the CANDLES/GOODS-N imaging
data. Extrapolations from lower redshift observations suggest that
these UV bright galaxies should be exceedingly rare (0.06 per {\em
  HST} field of view in the best case) by $z=11$. \bluetides, however,
predicts a significant probability of observing a galaxy like GN-$z11$
in the {\em HST} field of view ($\sim 13$ per cent). The \bluetides\
galaxies with $M_{\rm UV}\approx -22$ also match the inferred
properties of GN-$z11$ such as the age, mass, and star formation rate
\citep{Waters16}.

In this work, we use \bluetides\ to predict the properties of the
galaxy and active galactic nuclei (AGN) populations that will be
discovered by the \wfirst\ HLS and their clustering at $ 8 < z <15$.
In Section \ref{Section: BlueTides} we discuss the spectral synthesis
models we use to process the star formation, metallicity and stellar
ages from from \bluetides\ to determine galaxy luminosities. We
present a simple, self-consistent model for dust extinction, and also
find AGN luminosities. We present the \bluetides\ forecasts of the
photometric properties of galaxies detectable by the \wfirst\ HLS at
$z>8$ in Section \ref{Photometric Results}. In Section \ref{Section:
  Galaxy Properties} we present the \bluetides\ predictions for the
galaxy and AGN properties in the \wfirst\ HLS. We discuss the
clustering properties and bias of these galaxies in Section
\ref{Section: Correlation Functions} and conclude in Section
\ref{Section: Conclusion}.

\section{The BlueTides Simulation}
\label{Section: BlueTides}
With \bluetides\ (and the BlueWaters supercomputer at the National Center for Supercomputing Applications) a qualitative advance has been
possible: we have been able to run the first complete simulation (at
least in terms of the hydrodynamics and gravitational physics) of the
creation of the first galaxies and large-scale structures in the
universe. The application required essentially the full BlueWaters
system: we used 20,250 nodes (648,000 core equivalents).  In order to
effectively use the resources, a number of improvements had to be
applied to the cosmological code P-Gadget3 \citep{Springel} to make
P(eta)-Gadget, now MP-Gadget, which is now fully instrumented to run
on Peta-scale resources (F16). The major updates to the parallel
infrastructure allowed operation at BlueWaters scale. The simulation
code uses the pressure-entropy formulation of smoothed particle
hydrodynamics \citep{2013MNRAS.428.2840H} to solve the Euler
equations. The $(400h^{-1} \text{Mpc})^3$ cubic simulation volume
resulted in more than 200,000 star forming galaxies at redshift
$z=8$. Halos were identified using a Friends-of-Friends algorithm with
a linking length of 0.2 times the mean particle separation
\citep{Davis}. Table \ref{table:BlueTides Parameters} shows some of
the basic cosmological and computational values for \bluetides.

A variety of sub-grid physical processes were implemented to study
their effects on galaxy formation:
\begin{itemize}
\item star formation based on a multi-phase star formation model \citep{2003MNRAS.339..289S} with several effects following \citet{2013MNRAS.436.3031V}
\item gas cooling through radiative transfer \citep{1996ApJS..105...19K} and metal cooling \citep{2014MNRAS.444.1518V}
\item formation of molecular hydrogen and its effects on star formation \citep{2011ApJ...729...36K}
\item type II supernovae wind feedback (same model used in Illustris
  \citep{Illustris})
\item AGN feedback using the same model used in MassiveBlack I and II \citep{DiMatteo:2005ttp}
\item a "patchy" model of reionization \citep{2013ApJ...776...81B}
  with a fiducial model of mean reionization redshift $z\approx10$
  \citep{WMAP} with a UV background estimated by
  \citet{2009ApJ...703.1416F}
\end{itemize}

These sub-grid baryon-physics processes are important for determining
the photometric properties of high redshift galaxies and were
shown to produce results consistent with observations in F16. In this
section, we present the methods used to perform the post-processing of
\bluetides\ in order to forecast the photometric and clustering properties of
the galaxies detectable by \wfirst.

\subsection{Galaxy Luminosities and Dust Attenuation}
\label{Section: Luminosities}

Our methodology of producing synthetic galaxy photometry is detailed
in \citet{Wilkins}. In brief:
galaxy spectral energy distributions (SEDs) are calculated first by
assigning a pure-stellar SED to each star particle according to its
age and metallicity. We make use of the PEGASE v2 \citep{PEGASE}
stellar population synthesis (SPS) model assuming a \citet{Chabrier}
initial mass function (IMF). We note however that the choice of SPS
model can affect predicted UV luminosities by up to 0.1 dex
\citep{Wilkins}. Assuming a Salpeter IMF, instead of a
\citet{Chabrier} IMF, results in luminosities approximately 0.2 dex
lower.

To model dust attenuation we utilize a scheme inspired
by \cite{sunrise2006}, where the metal density is integrated along
parallel lines of sight. The dust attenuation is assumed to be
proportional to the surface density of metal along the line of sight,
\begin{equation}
\tau_V(x, y, z) = \kappa \Sigma (x, y, z) = \int_{0}^{z} dz'\kappa \rho_\mathrm{metal}(x, y, z')
\label{dust_attenuation}
\end{equation}
where $\tau_V(x, y, z)$ is the dust optical depth,
$\rho_\mathrm{metal}(x, y, z')$ is the metal density, $\kappa$
is a normalization factor, and we have
chosen the $z$ direction as the the line of sight direction. The dust
attenuation is then applied at subsequent redshifts 
(for more details see Wilkins et al. {\em in-prep}).

This simple model is calibrated so that we reproduce the bright end of
the $z\sim 8$ UV LF from B15 (we naturally reproduce the faint end
where we predict very little attenuation). It is however important to
note that simply linking the metal density to the dust optical depth
may not fully capture the redshift and luminosity/mass dependence of
dust attenuation. This is because the production of dust is not
expected to fully trace the production of metals \citep[e.g.][]{2015MNRAS.451L..70M} with the expectation of lower
dust-to-metal ratios at earlier times, and thus lower attenuation at
high redshift. This is supported by the recent discovery of GN$-z11$
\citep{Oesch16, Waters16}. In the rest of this work, we
present the predictions for both the dust corrected and intrinsic LF
of galaxies in \bluetides\ for all redshifts.

\subsection{Active Galactic Nuclei}

Supermassive black holes were seeded in \bluetides\ with an initial
mass of $5\times10^5h^{-1}M_\odot$ once a halo reached a mass greater
than $5\times10^{10}h^{-1}M_\odot$. \bluetides\ tracks the rate at
which mass is accreted from the halo onto the super-massive blackhole
and from that a bolometric luminosity can be found. The blackhole
bolometric luminosity is computed by assuming a mass-to-light
conversion efficiency $\eta = 0.1$:
\begin{eqnarray}
L = \eta c^2\frac{dM_{\rm BH}}{dt}
\label{AGN_L}
\end{eqnarray}
where $L$ is the bolometric luminosity and $M_{\rm BH}$ is the mass of the
black hole. Black holes in the \bluetides\ simulation were limited to
having a mass accretion rate of three times that of their Eddington
limit, 
given by $\dot{M}_\mathrm{edd}=4\pi G c M_{\rm BH} m_p/\sigma_T$ where $G$,
$c$, $m_p$, and $\sigma_T$ are Newton's constant, the speed of
light, the mass of the proton, and the
Thomson scattering cross section, respectively. 
The UV magnitude of the AGN is determined by \citep{Fontanot}:
\begin{equation}
M_{\rm UV} = -2.5\log_{10}\frac{L}{f_{\rm bol}\nu_{\rm B}}+34.1+\Delta_{\rm B,UV}
\end{equation}
where $f_{\rm bol}=10.2$ is the bolometric correction \citep{Elvis}, $\nu_{\rm B}=6.74\times10^{14}{\rm Hz}$, and $\Delta_{\rm B,UV}=-0.48$. 

\subsection{Clustering}
\label{Clustering Methods}
Utilizing the spatial positions of galaxies in the \bluetides\
simulation, we compute the two-point spatial correlation function
$\xi_\mathrm{gg}(r)$. 
For a particular redshift, we use all galaxies brighter than the 
\wfirst\ magnitude limit in the periodic volume of the box
and use direct pair counts to measure $\xi_\mathrm{gg}(r)$. 
We also compute the dark matter correlation function,
$\xi_\mathrm{dm}(r)$
 using a
Fast Fourier Transform based algorithm
\footnote{\url{https://github.com/bccp/nbodykit}}.

For each redshift we 
compute the linear bias defined by
\begin{equation}
b = \sqrt{\frac{\xi_\mathrm{gg}(r)}{\xi_\mathrm{dm}(r)}}
\label{bias}
\end{equation}
 We
compute $b$ by averaging
equation \ref{bias} between the distances $r=10h^{-1} \mathrm{Mpc}$
and $r=75h^{-1} \mathrm{Mpc}$ to stay in the linear regime.

\section{The WFIRST HLS Galaxy Population}
\label{Photometric Results}

\subsection{Predicted Luminosity Functions}

\begin{figure}
\begin{minipage}[c][][t]{0.45\textwidth}
  \vspace*{\fill}
  \centering
  \includegraphics[width=\textwidth]{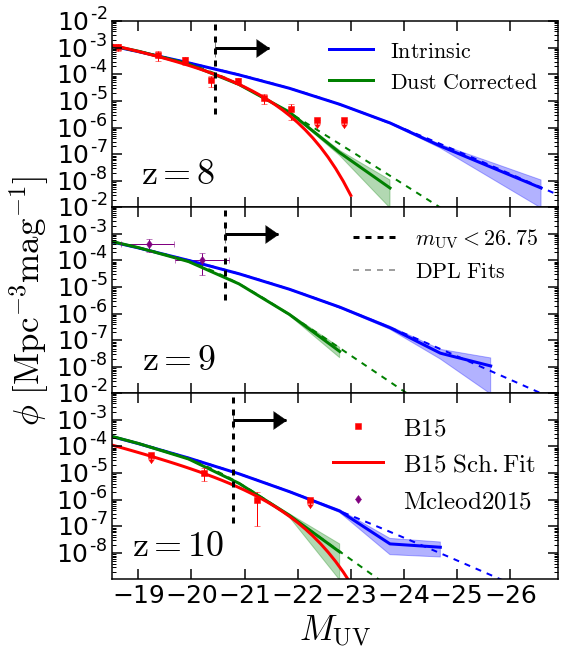}
  \par\vfill
  \includegraphics[width=\textwidth]{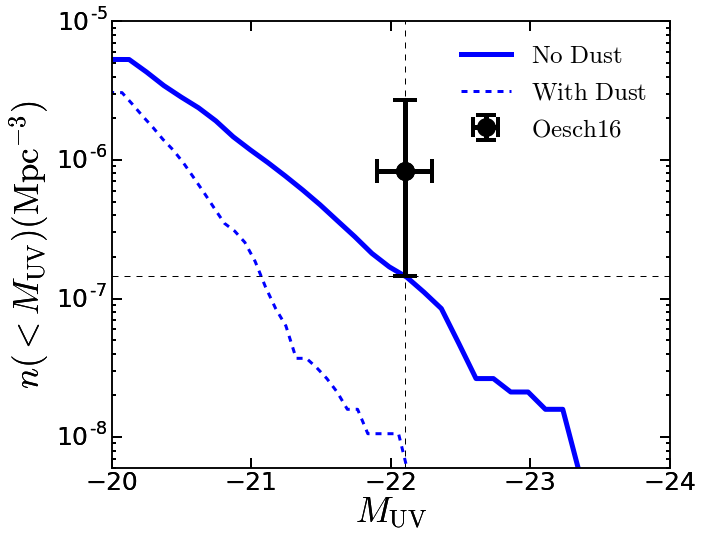}
  
\end{minipage}
\caption{Top Panel: Comparison of the intrinsic (solid blue lines) and
  dust corrected (solid green lines) \bluetides\ LFs to available
  observational data from B15 (red data points) and \citet{McLeod} (purple data points). Shaded regions show the $1\sigma$ Poisson errors. Also shown are the
  best DPL fits (dashed blue and green lines) to the \bluetides\ LFs and the
  Schechter fits to observation data from B15 (red lines). The galaxies to the right of the dashed black lines
  are those that are UV bright enough to be detected by
  \wfirst. Bottom Panel: \bluetides\ prediction for the number density
  of objects like GN-$z11$. The error bar on the \citet{Oesch16} point is
  a simple Poisson error.}
    \label{fig:obs_comp}
\end{figure}

The intrinsic and dust corrected LFs from the galaxies in \bluetides\
are shown in Figure \ref{fig:obs_comp} for $z=8$, 9 and 10, with Poisson errors given by the shaded regions.  The
dashed blue and green lines show a double power law (DPL) fit to the
\bluetides\ intrinsic and dust corrected LFs, respectively. We note
that a standard Schechter function does not represent a good
description of the UV LF in \bluetides\ as it consistently under
predicts the bright end (with and without accounting for dust). Here
we use a DPL \citep[B15;][]{Bowler14} defined by:
\begin{equation}
\phi(M) = \frac{\phi^*}{10^{0.4(\alpha+1)(M-M^*)}+ 10^{0.4(\beta+1)(M-M^*)}}
\label{dpl}
\end{equation}
where $\phi^*, M^*, \alpha, {\rm and\ } \beta$ are the normalization,
characteristic magnitude, faint end slope, and bright end slope, 
respectively. Results and the parameters of the DPL fits are presented
in the Appendix.

The data points in Figure \ref{fig:obs_comp} show the observations
compiled by B15 and \citet{McLeod}. The red line is the best fit
Schechter function as provided by B15. The arrows in the
left-hand-side of the plot in Figure \ref{fig:obs_comp} indicate the
planned $5\sigma$ detection limit of \wfirst\ HLS. We note that the
current observational data shows no deviation from a Schechter-like LF
at these high redshifts (see also discussion in B15). However, the
bright end ($M_{\rm UV} < -21$) of the UV LF at $z>8$ is largely
unconstrained observationally. \bluetides\ predicts a deviation from a
Schechter LF in the \wfirst\ galaxy population. In particular, the
number of objects predicted for the bright end of the LF is
significantly enhanced compared to those expected from the
extrapolation of the Schechter fit to current observational data.

\begin{figure}
\includegraphics[width = 0.45\textwidth]{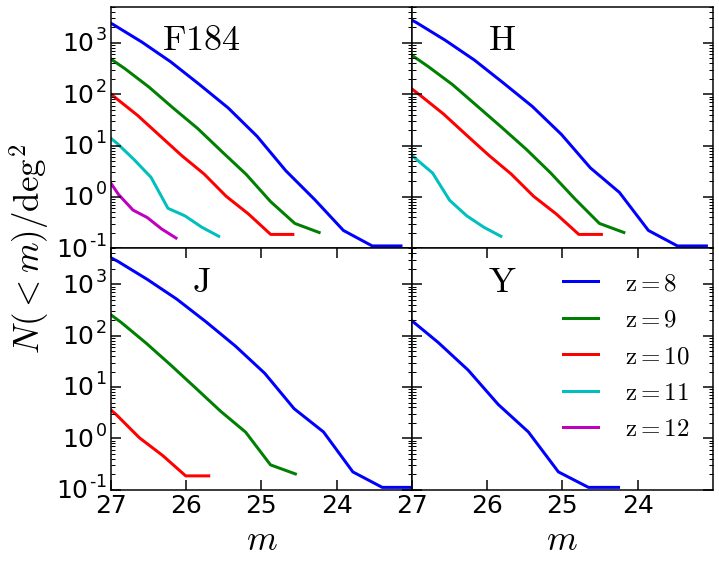}
\caption{Surface density photometry for the \wfirst\ HLS band
  passes. Luminosities bey ond the Lyman break limit have been set to
  zero for the appropriate redshift in each band.}
\label{fig: galaxy_photometry}
\end{figure}

The \bluetides\ prediction of an enhanced number of bright sources is
supported by the recent discovery by \citet{Oesch16} of the $M_{\rm
  UV} = -22.1$ galaxy at $z=11$ \citep[see][]{Waters16}. In the bottom
panel of Fig~\ref{fig:obs_comp} we show the number density inferred by
\cite{Oesch16} at $z=11$ compared to the \bluetides\ prediction (with
and without dust). The \bluetides\ predictions using the intrinsic LFs
is consistent with the observation of this incredibly massive $z=11$
galaxy \citep{Waters16}. Note also that our dust correction model
reduces the number density of objects like GN-z11 by an order of
magnitude, making it largely incompatible with the observation of
\citet{Oesch16} and likely implying an increasingly negligible amount
of dust at these redshifts (see also Section \ref{Section:
  Luminosities}).
  
\begin{figure}
\begin{minipage}[c][][t]{0.45\textwidth}
  \vspace*{\fill}
  \centering
  \includegraphics[width=\textwidth]{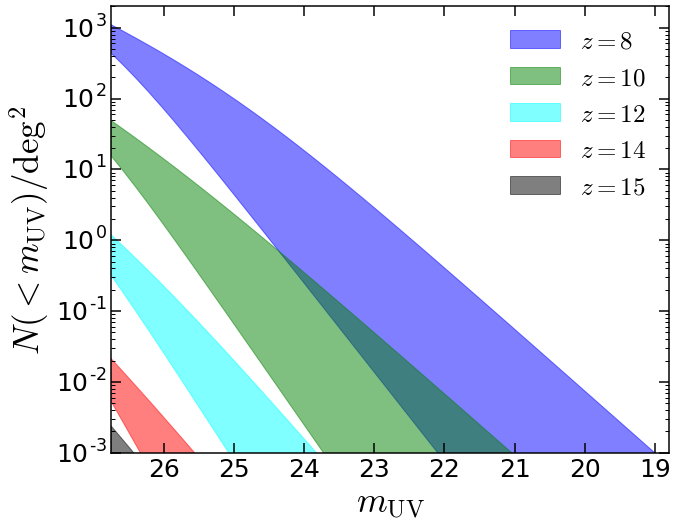}
  \par\vfill
  \includegraphics[width=\textwidth]{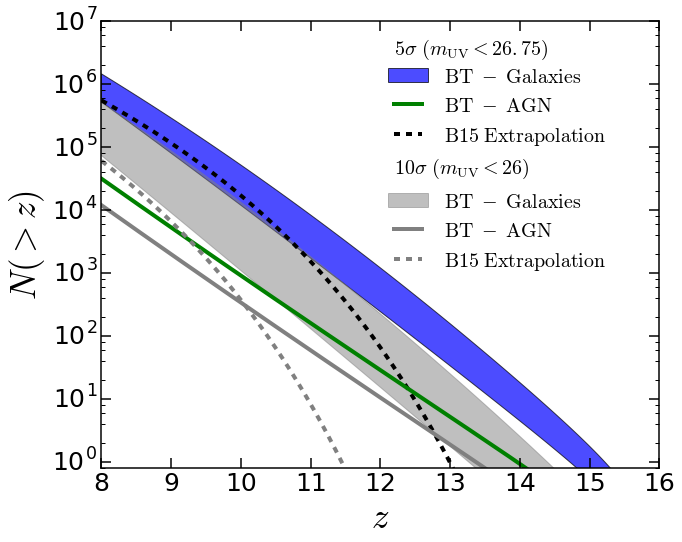}
  
\end{minipage}
\caption{Top Panel: Surface density for $z=8$ and beyond. The results
  for $z=9,11,$ and $13$ are left out for clarity. The \wfirst\ field
  of view is $\sim 2200 {\rm deg}^2$, so will be able to detect
  galaxies out to $z=15$ according to the \bluetides\ LF results. Our
  dust model is too extreme by $z=10$, so the lower limit at these
  redshifts are very conservative estimates. 
  Bottom Panel: Predicted
  number of galaxies in the \wfirst\ HLS at or above a given redshift. The predictions shown are for the \bluetides\ galaxies (blue region) and AGN (green line) for the $5\sigma$ limit. Extrapolation of the B15 $z=8$ Schecther function is also shown (black dashed line). The grey lines show the same quantities but for the $10\sigma$ limit.}
 \label{fig:N_v_m}

\end{figure}

\subsection{Expected Number of Galaxies in the HLS}
\label{Section: Expected Number}

Given the calculted SEDs for each galaxy in \bluetides\, Figure
\ref{fig: galaxy_photometry} shows the predicted photometry (without
dust corrections) for the F184(1.683-2.000 \micron), H (1.380-1.774
\micron), J (1.131-1.454 \micron), and Y (0.927-1.192 \micron) bands
which are used in the \wfirst\ HLS for galaxy selection (S13) in the
\bluetides\ volume. For redshifts $z=9-12$ we have set the luminosities
at wavelengths beyond the Lyman break to zero to account for
absorption by the intergalactic medium.

In order to make predictions for the total number of galaxies in the
HLS we use the the DPL fits to the \bluetides\ LFs to extrapolate to
the bright end. This is necessary as The \wfirst\ HLS has a much
larger observational volume than the \bluetides\ simulation cube
(approximately by a factor of 400). The number of galaxies per
unit solid angle between $z_1$ and $z_2$ brighter than absolute
magnitude $M$ is simply given by
\begin{equation}
  N =\int_{z_1}^{z_2}dz\ r^2\frac{dr}{dz}\int_{-\infty}^M dM' \phi(M',z)\ sr^{-1}
\label{cum_num_eqn}
\end{equation}
where $r=r(z)$ is the comoving distance to redshift $z$.  In the top
panel of Figure \ref{fig:N_v_m} we show the expected surface densities
in the \wfirst\ HLS as a function of the rest-frame UV luminosities
(for all magnitudes less than the \wfirst\ $5\sigma$ limit). We
calculate the surface density at each redshift $z$ with equation
\ref{cum_num_eqn} between redshift $z-0.5$ and $z+0.5$ as a function
of apparent magnitude. The upper (lower) limits at each redshift are
from the LF without (with) dust correction. We give redshift evolution
fits for the number densities in the Appendix.

The bottom panel of Figure \ref{fig:N_v_m} shows the total cumulative
number of galaxies above a given redshift for the $5\sigma$ and
$10\sigma$ \wfirst\ limits. We calculate the cumulative number for
each redshift using equation \ref{cum_num_eqn} with a magnitude limit
at each redshift that corresponds to the appropriate \wfirst\
limit. For comparison we also show the results for the cumulative
number obtained using the B15 $z=8$ Schechter function best fit
considering an optimistic evolution
($dM^*/dz=0.36$ as in S13) of the characteristic magnitude. The pessimistic evolution from S13 ($dM^*/dz=1.06$) significantly under fits the $z=10$ results from B15 and are therefore not considered. The
\bluetides\ LFs predicts a comparable total number of galaxies as the
B15 extrapolation but \bluetides\ predicts objects out to a
significantly higher redshift. This is a result of the near identical
normalizations of the \bluetides\ and B15 $z=8$ LFs, but an enhanced
bright end in \bluetides\ compared to the extrapolation of observed
constraints.
\bluetides\ predicts a total of $10^6$ galaxies beyond $z=8$, and up
to a few at $z=14-15$ at the planned depth and area of the \wfirst\
HLS.  Without much dust, \wfirst\ will detect galaxies out to $z=15$
at the $5\sigma$ level.

\begin{figure}
\includegraphics[width=0.45\textwidth]{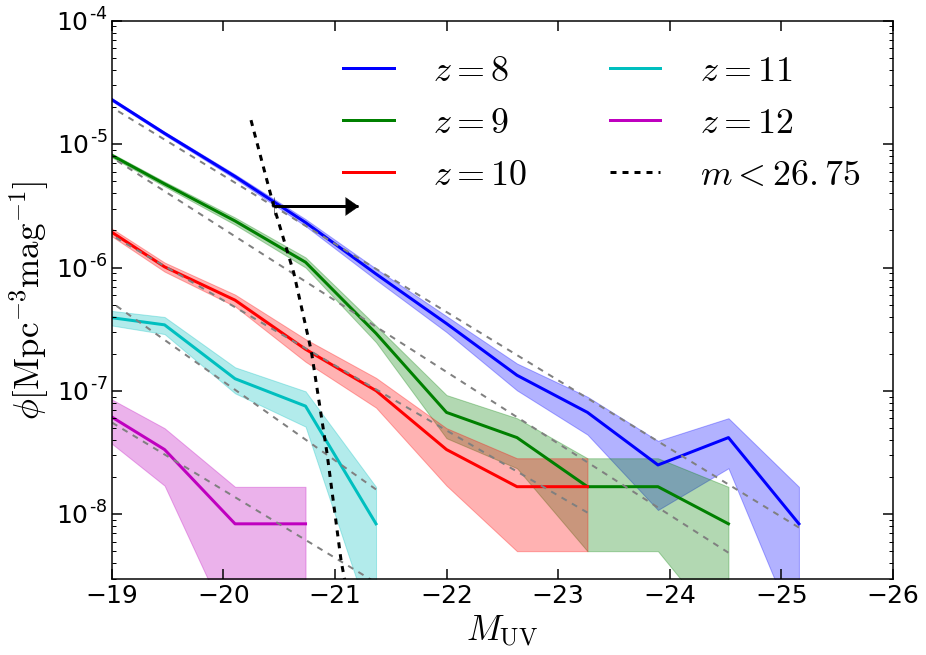}
\caption{AGN LFs for \bluetides. Shaded regions show the $1\sigma$ Poisson errors.  Grey dashed lines show a best fit to a power law LF. The black arrow and dashed line indicates the galaxies that are UV brighter than the \wfirst\ $5\sigma$ limit.}
\label{fig:agn_lf}
\end{figure}

\begin{figure}
\includegraphics[width=0.45\textwidth]{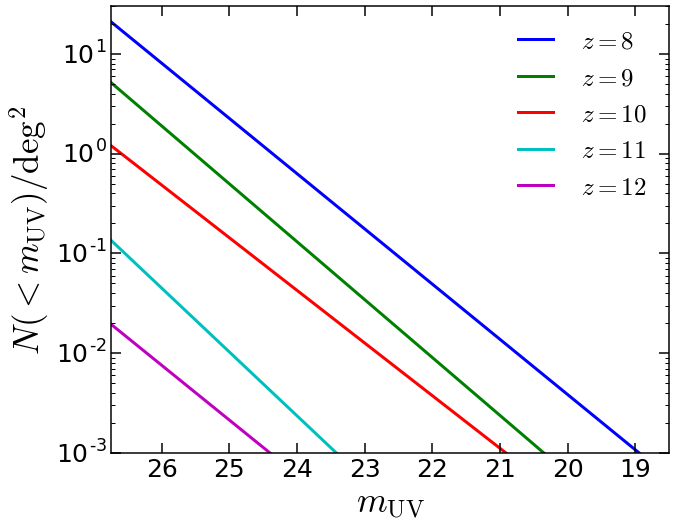}
\caption{AGN surface densities according to \bluetides\ for AGN that have an intrinsic brightness above the \wfirst\ $5\sigma$ limit (no dust correction assumed).}
\label{fig:bh_N}
\end{figure}

\subsection{Expected Number of AGN}
The proposed survey area and depth of the \wfirst\ HLS will allow for
the discovery of substantial populations of high-$z$ AGN.  Currently
the highest redshift known quasar is at $z=7.1$ \citep{Mortlock} and only a handful of objects known at $z > 6$ from SDSS
\citep{Fan:2005es}. The limited knowledge of high-$z$ black
holes will be revolutionized by \wfirst\ HLS. Using the black hole
population simulated in \bluetides\ we examine the predictions for
the LF and the expected number of AGN in the HLS at $z\ge 8$.  Figure
\ref{fig:agn_lf} shows the intrinsic UV LF for AGN in \bluetides\ with
the dashed black line indicating the detection limit of \wfirst\ HLS. 
In order to predict the total number of AGN expected in the field of
the HLS we fit the AGN LFs to a power law which we can then
extrapolate to obtain the surface density of AGN above a given
magnitude for those AGN beyond the $5\sigma$ cutoff for \wfirst\ (Figure \ref{fig:bh_N}). Figure \ref{fig:N_v_m} (green and grey lines)
shows tens of thousand of AGN could be detectable at the $5\sigma$
level, with the brightest AGN  out to $z=13-14$.

\section{WFIRST Galaxy and AGN Properties}
\label{Section: Galaxy Properties}

In this section, we examine a number of fundamental properties of the
galaxy and AGN population predicted in the \wfirst\ HLS by \bluetides.

\subsection{Galaxy Properties}

\begin{figure}
\includegraphics[width = 0.45\textwidth]{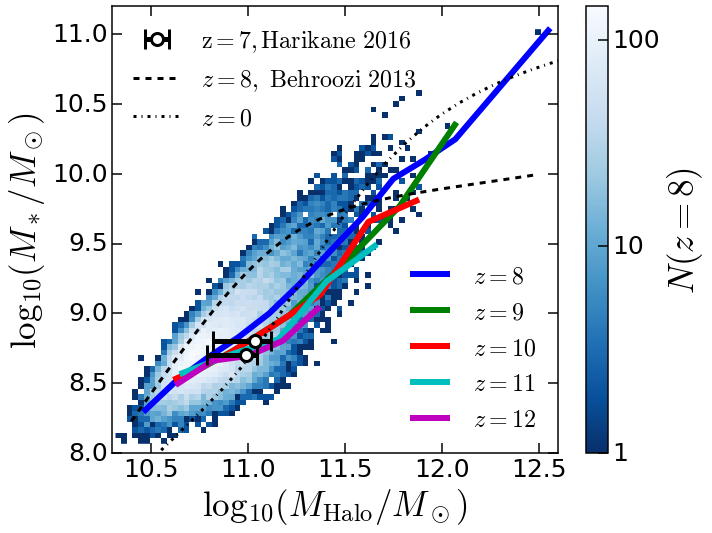}
\caption{Stellar mass vs halo mass for the \wfirst\ galaxy population
  as predicted by \bluetides. Dashed lines show results from abundance
  matching by \citet{Behroozi}. Data points
  show the $z=7$ results from \citet{Harikane2015}. The 2D histogram
  shows the distribution of galaxies in the \bluetides\ volume at
  $z=8$. Solid lines show the mean in bins of $M_{\rm Halo}$ for the
  higher redshifts as indicated in the figure.}
\label{fig:sm_hm}
\end{figure}

In Figure \ref{fig:sm_hm}, we show the stellar - halo mass relation
for the galaxies in \bluetides. Galaxies detected in our mock HLS have
stellar masses between $10^8-10^{10}M_\odot$ that are hosted by dark
matter halos with masses of $10^{10.5}-10^{12}M_\odot$ which are
correlated with each other. The mean is well fit by a power law at
each redshift, $M^* = (M_h/M_0)^\alpha$ where the slope shows mild
redshift evolution ($\alpha\approx-0.2(1+z) + 2.8$).  We compare the
relation in \bluetides\ with the abundance matching results from
\cite{Behroozi} at $z=0$ and extrapolated to $z=8$. The galaxy
population in \bluetides\ appears somewhat closer to $z=0$ relation but
perhaps showing less prominent signs of quenching at the high mass
end. It is interesting to note that the predicted relation is in
agreement with the halo occupation distribution modeling results \citep{Harikane2015}. The $z=7$ stellar - halo mass relation of \citet{Harikane2015} comes from the observations of around 300 Lymann break galaxies in the GOODS-N and GOODS-S fields. They split their $z=7$ sample into two subsamples ($m_{\rm UV} < 28.2$ and $m_{\rm UV} <28.4$) each with around 100 galaxies and occupy halos with a model that assumes the number of galaxies in a given halo only depends on the halo mass.

\begin{figure}
\includegraphics[width = 0.45\textwidth]{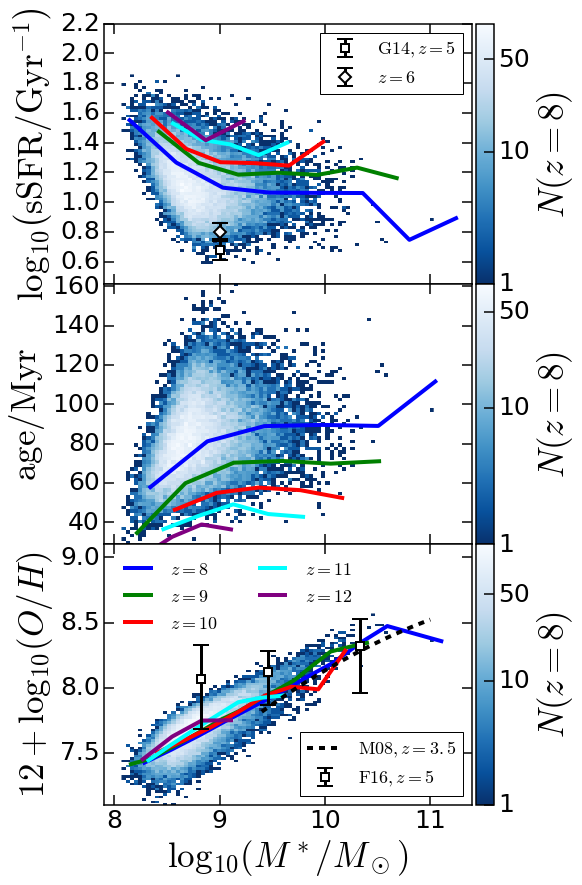}
\caption{Galaxy properties for those bright enough to be detected by
  \wfirst. The 2D histogram shows the distribution of galaxies in the
  \bluetides\ volume at $z=8$. The solid lines show the mean values in bins of
  stellar mass. sSFR observations come from \citet{Gonzalez2014}. Best fit to metallicity measurements at $z=3.5$ come from \citet{Maiolino2008} and observations at $z=5$ come from \citet{Faisst2016}.}
\label{fig:gal_prop}
\end{figure}

In Figure \ref{fig:gal_prop} we show various galaxy properties from
$z=8$ to $12$. The top panel shows the specific star formation rate
(SFR$/M^*$) as a function of stellar mass. SFR$/M^*$ shows a strong
evolution with redshift with a dependence ${\rm sSFR} \approx [4.1(1+z)
- 24.5]\ {\rm Gyr}^{-1}$ at $M^* = 10^9 M_{\odot}$. 
The \bluetides\ specific star formation rate compares well 
with the redshift evolution trend observed in the GOODS-S field of view \citep{Gonzalez2014}.

The age of galaxies (defined to be the mean time since formation of
their constituent star particles) range from $\sim 80$ Myr at $z=8$ to
$\sim 40$ Myrs old at $z=12$. We find (for $M^* = 10^9 M_\odot$) a
redshift dependence such that ${\rm age} \approx [-11.1(1+z)+180]\
{\rm Myr}$.

Gas-phase metallicity as a function of stellar mass (mass-metallicity,
MS, relation) is shown in the bottom panel of Figure
\ref{fig:gal_prop}, where
$\log_{10}(Z/Z_\odot)+8.69=12+\log_{10}(O/H)$ \citep{Asplund2009} and
we have assumed $Z_\odot = 0.02$.  We define the metallicity using the
star forming star particles within the galaxies, which are typically
centrally concentrated. This is likely to most closely match with how
metallicities are measured observationally from star-forming regions.
We point out that a different definition of the gas-phase metallicity
that includes all gas particles (not only SF) in galaxies leads to
values about 0.5 dex lower than what shown in Fig.~\ref{fig:gal_prop}.
The mass-metallicity relation for the mock HLS galaxies shows
negligible redshift evolution and a dependence on stellar mass
($\frac{d\log_{10}Z}{d\log_{10}M^*}\sim0.4$ with $M^*$ in units of
$M_\odot$). The slope of the predicted MS relation is consistent to
the observed one. Recent measurements of the MS relation at high-z has
been carried out by \citet{Faisst2016} who find comparable
metallicities for $z\sim5$ in the COSMOS fields to the $z=3.5$ results
of \citet{Maiolino2008}, implying a weak dependence with redshift for
$z > 3.5$.  The \bluetides\ results predict an amplitude consistent
with the measurements of \citet{Faisst2016} and \citet{Maiolino2008},
and supports the lack of redshift evolution of metallicity at
high-z. However, the measurements of the MS relation at these high
redshifts still has large uncertainties and even at low redshfits (not
shown here), these measurements are subject to calibration issues that
can change values up to $0.7{\rm\ dex}$ \citep{Kewley2008}. This makes
it hard to place constraints on \bluetides\ star formation models and
feedback processes based on predictions for the amplitude of the MS
relation.

\begin{figure}
\includegraphics[width = 0.45\textwidth]{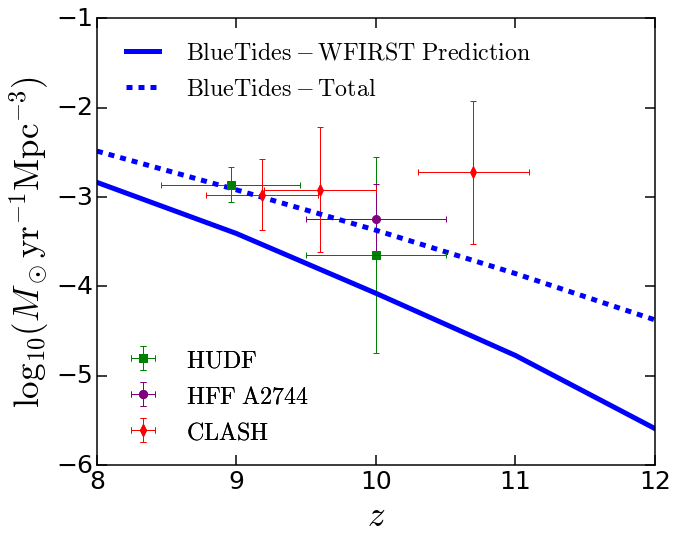}
\caption{Star formation rate density prediction for the \wfirst\
  HLS. The solid line shows the predicted result for \wfirst, without
  dust correction. The dashed line shows the contribution from the
  total \bluetides\ galaxy sample. Observational data points are for the
  HUDF \citep{Oesch14b}, HFF A2744 \citep{Oesch14a}, and observational
  estimates from CLASH \citep{Bouwens14b, Zheng12, Coe13}}
\label{fig:ssfr}
\end{figure}

Figure \ref{fig:ssfr} shows the global star formation rate density
prediction from \bluetides\ for galaxies in the \wfirst\ HLS, together with
current  observational constraints.  It is important to note that the
observational results are from galaxies with $M_{UV} < -18$ whereas at
$z=8$, the \wfirst\ $5\sigma$ limit corresponds to $M_{UV} <
-20.5$. This shows that galaxies fainter than this magnitude
 contribute $\sim55$
per cent of the observed star formation rate density at $z=8$. By
$z=12$, \wfirst\ will only be able to directly probe $\sim6$ per cent
of the total star formation rate density.

\subsection{AGN Properties}
\label{Section: AGN Properties}

\begin{figure}
\includegraphics[width=0.45\textwidth]{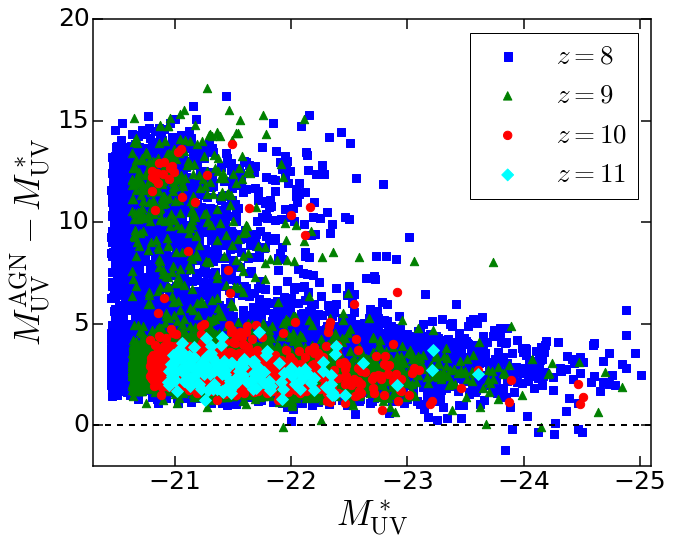}
\caption{Magnitude difference of AGN and their host galaxies. Points below the dashed black line indicate AGN that out shine their host galaxy.}
\label{fig:agn_cont}
\end{figure}

Figure \ref{fig:agn_cont} shows the relative brightness of AGN and
their host galaxies for the \wfirst\ galaxy population. Only about 0.3
per cent of the AGN outshine their host galaxies at $z=8$ and $9$,
while the rest of the AGN are $\gtrsim 1$ magnitude fainter than their
host galaxy.  These exceptionally bright black holes are those with
the highest mass (see Di Matteo et al. (2016) {\em in prep}).  By
$z=10$, all of the AGN in the \bluetides\ volume are $ \gtrsim1$
magnitude fainter than their host galaxy.

\begin{figure}
\includegraphics[width=0.45\textwidth]{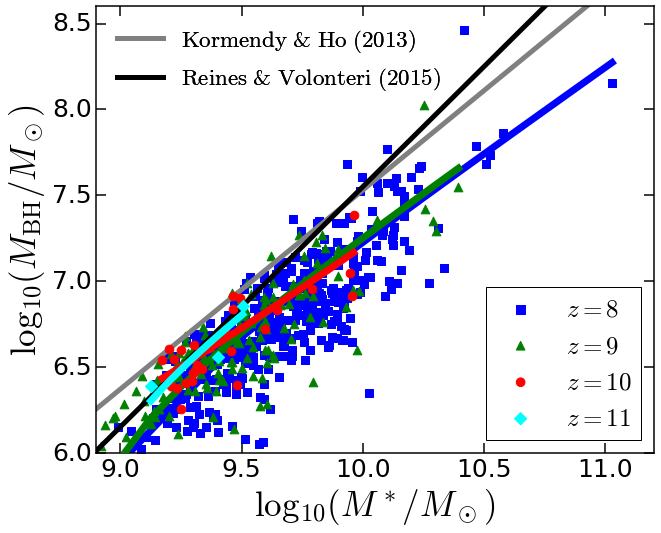}
\caption{Black hole mass vs. stellar mass for the AGN brighter than the \wfirst\ $5\sigma$ limit. Data points show the distribution at each redshift. The solid lines show the best linear fit at each redshift. Grey and black lines show the best fits to local observations ($z<1$) from \citet{KormendyHo2013} and \citet{Reines2015}, respectively.}
\label{fig:sm_bhm}
\end{figure}

\begin{figure}
\includegraphics[width=0.45\textwidth]{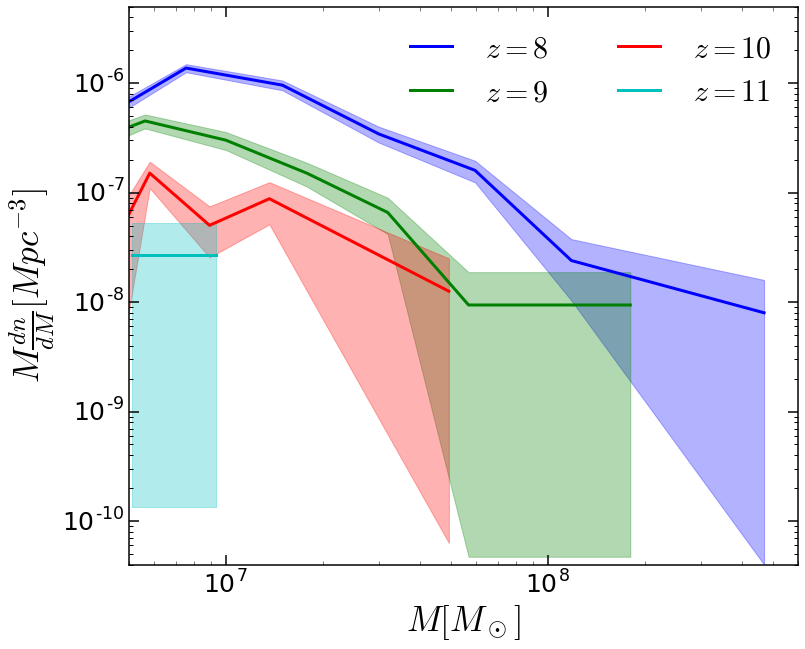}
\caption{Black hole mass function for the AGN with intrinsic luminosities bright enough to be detected by \wfirst. Shaded region indicates $1\sigma$ Poisson errors.}
\label{fig:bh_mf}
\end{figure}

\begin{figure}
\includegraphics[width=0.45\textwidth]{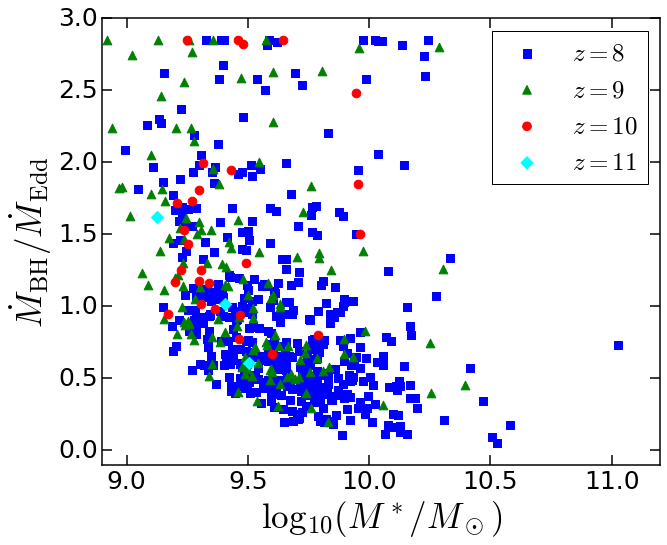}
\caption{AGN mass accretion rate vs. stellar mass for the AGN with intrinsic luminosities bright enough to be detected by \wfirst.}
\label{fig:bh_marf}
\end{figure}

Figure \ref{fig:sm_bhm} shows black hole mass as a function of stellar
mass, as well as a linear fit at each redshift. We find that for the
AGN detectable by the \wfirst\ HLS, the $M_{\rm BH}/M^*$ ratio is
about $2\times10^{-3}$ at all redshifts, showing little redshift
evolution. We show a comparison to local observations ($z<1$) from
\citet{KormendyHo2013} and \citet{Reines2015}. Our high redshift
results indicate a similar $M_{\rm BH}/M^*$ ratio to the local
universe, but with about a $0.2$ dex offset in normalization. Figure
\ref{fig:bh_mf} shows the black hole mass function (number density of
black holes per unit mass) for the AGN brighter than the \wfirst\
$5\sigma$ cutoff. We see that the central super-massive black holes
range in mass from $10^6-10^8 h^{-1}M_\odot$. Figure \ref{fig:bh_marf}
shows the ratio of the black hole mass accretion rate to its Eddington
rate as a function of stellar mass. These AGN are accreting at a rate
$\gtrsim0.5\dot{M}_\mathrm{edd}$.

\begin{figure*}
\centering
\includegraphics[width = 0.95\textwidth]{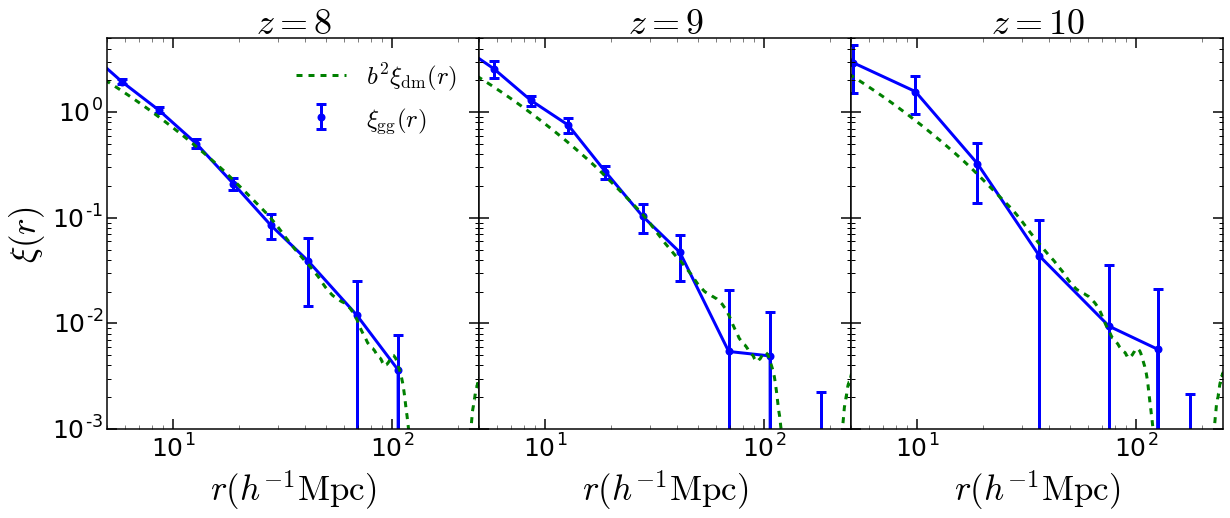}
\caption{Galaxy correlation functions for redshifts $z=8,9,$ and $10$. Error
  bars are delete-one jacknife errors. Green dashed line shows the dark matter
  correlation function multiplied by the square of the linear bias.}
\label{fig: cf_all_z}
\end{figure*}

\begin{figure}
\includegraphics[width = 0.45\textwidth]{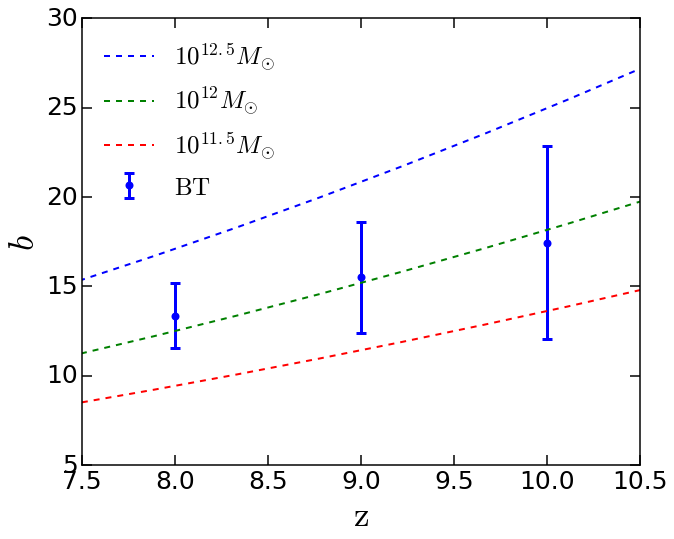}
\caption{Bias measurements for the galaxies visible to \wfirst\ from \bluetides\ (blue points). Also shown is the linear bias computed from \citet{Tinker} for a given halo mass (dashed lines).}
\label{fig:bias}
\end{figure}

\section{Clustering}
\label{Section: Correlation Functions}
Given the large sample of galaxies expected in the \wfirst\ HLS, it will
be possible to measure their correlation functions at these high redshifts.
Measurement of three-dimensional clustering will require 
galaxy redshift information. The \wfirst\ Grism survey is planned to cover 
about one third of the area of the HLS. It is designed primarily
 for searching for 
emission line galaxies with redshifts $z\approx1-3$, using their H$\alpha$ lines 
and the OIII doublet. As a result the wavelength coverage is planned to be 
1.35-1.89 microns with R=461.
The galaxy described by \citet{Oesch16} was confirmed to be at redshift
$z=11.09$ from observation of the Ly$\alpha$ break in {\em HST} Grism spectroscopy. 
Given that we predict from \bluetides\ that large populations of galaxies 
of similar apparent magnitude exist at redshifts $z=8$, it is possible
that redshifts can be measured to similar accuracy, depending on the 
performance of \wfirst\ spectroscopy. One obstacle to measuring
redshifts below $z=10.1$ however is that the current lower limit of the
\wfirst\ Grism coverage would exclude detection of the  Ly$\alpha$ break.

Without spectroscopy, multiband photometric redshifts will be  obtained
for the galaxies we investigate in this work. For example, \citet{Calvi2016} have recently extended the  BORG (Brightest of Reionizing
Galaxies) survey to $z=9-10$ using {\em HST} five-band photometry. They find
that approximately $30$ per cent of galaxies are interlopers with $z\sim1$, whose 
Balmer break is masquerading as a Ly$\alpha$ break at higher z. 
This contamination results in a suppression in angular clustering which should
be modeled. In this work, we concentrate on three dimensional
measurements of the correlation function to illustrate the clustering bias and
other properties of the galaxies in \bluetides. We leave
specific modeling of the impact of Grism redshifts on three 
dimensional clustering (or projected clustering), and also angular clustering  
(for example using photometric information only) to future work. As a result,
our speculations below on the possibility of detecting the BAO feature in
clustering represent a best case scenario, intended to spur research
into the idea of making a measurement at these high redshifts.

We calculate the correlation functions from the \bluetides\ galaxies as 
described in Section \ref{Clustering Methods}. 
We compute the correlation functions in real space, as we expect the 
redshift distortions of these highly biased objects to be small.
To estimate errors on the 
galaxy correlation function, we utilize a delete-one jacknife method. 
We divide the \bluetides\ volume into eight sub samples and calculate the 
correlation function, removing each sub sample one at a time. 
The errors on the correlation function are then given by
\begin{equation}
\sigma^2(r) = \frac{N - 1}{N} \sum_i^N (\xi(r) - \xi_i(r))^2
\end{equation}
where $\xi(r)$ is the correlation function for the entire volume and
$\xi_i(r)$ is the correlation function with the $i$th sub volume
removed. Figure \ref{fig: cf_all_z} shows the measured correlation function of
the \bluetides\  galaxies brighter than the \wfirst\ $5\sigma$ limit. 
Also shown is the dark matter correlation function. 
We multiply the dark matter  correlation function by the bias computed in
equation \ref{bias}. 

\citet{Barone-Nugent:2014eqa} measured a linear bias of $b=8.6$ using about 650 Lyman Break Galaxies at $z\sim 7$ in the {\em HST} Ultra Deep Fields. The \wfirst\ HLS will be capable of detecting $\sim 650$ galaxies at $z=13$ according to \bluetides. So \wfirst\ may have 
a large enough galaxy sample to determine the linear bias out to $z\sim13$.
The galaxy bias as a function of redshift is shown in Figure \ref{fig:bias} 
with errors propagated forward from the jacknife errors on the 
galaxy correlation functions (no error assumed on the dark matter only 
correlation function). The \wfirst\ galaxy population in \bluetides\ implies
values of $b=13.4\pm1.8$ at $z=8$, 
which will be the largest bias ever measured. The galaxy bias increases
linearly with redshift ($b\approx 2.1(1+z) -5.3$)
reaching values close to $b\sim 20$ by $z=11$. 

Figure \ref{fig: cf_all_z} shows that at $r \sim 100h^{-1}$Mpc, the
\bluetides\ galaxy correlation function is consistent with no BAO peak. This
is a result of the relatively small simulation volume of \bluetides. 
However \wfirst\ HLS will cover a much bigger comoving volume (by a factor of $\sim 400$). 
The detectability of the BAO peak is largely determined by the survey volume and the product
$nP_{\rm gg}$, where $n$ is the number density of the galaxy population and
$P_{\rm gg}= b^2 P_{\rm dm}$ is the galaxy power spectrum evaluated at the BAO
scale $k=0.2h \rm Mpc^{-1}$.   
 For $nP_{\rm gg}<1$, the error on the measurement of the BAO peak is dominated by shot noise. For $nP_{\rm gg}>1$ the BAO peak can be fully sampled for a strong detection. Given the bias measured above and the number
density of galaxies in \bluetides\ (and
the dark matter only power spectrum from \citet{Eisenstein1998} for $P_{\rm
  dm}$) we can  estimate the expected strength of the BAO signal. 

\begin{table}
\centering
\begin{tabular}{||c c c ||} 
 \hline
 & BOSS CMASS & \wfirst\  \\ [0.1ex] 
 \hline\hline

$\bar{z}$ & 0.57 & 8.0 \\
$ z$ & 0.43 < z < 0.7 & 7.865 < z < 8.135 \\
$P_\text{dm} (h^{-3} Mpc^3)$ & 2040 & 75.3 \\
$b$ & 2.0 & 13.36 \\
$P_\text{gg} (h^{-3} Mpc^3)$ & 8160 & 13240 \\
$n\ (h^{3}Mpc^{-3})$ & $3.0\times10^{-4}$ & $1.19-3.6\times10^{-4}$ \\
$nP_\text{gg}$ & {\bf 2.45} & {\bf 1.60-4.87} \\
$A_\text{Survey}\ (\text{deg}^2)$ & 3275 & 2200 \\
$V\ (Gpc^3)$ & {\bf 3.89} & {\bf 4.58}\\

 \hline
\end{tabular}
\caption{Comparison of the BOSS CMASS sample \citep{Anderson:2012} and the \wfirst\ $z=8$ sample as predicted by \bluetides. $P_{\rm gg}$ and $P_{\rm dm}$ are computed at the BAO scale $k=0.2 h/Mpc$ at the mean redshift of the sample in question. Ranges for $n$ (and $nP$) are computed using galaxies brighter than the \wfirst\ $5\sigma$ according to the intrinsic and dust corrected galaxy luminosities. $V$ is the comoving volume for each survey's respective area coverage, assuming the same range in redshift centered about the mean redshift $\bar{z}$.} 
\label{table:boss_wfirst_comp}
\end{table}

Table \ref{table:boss_wfirst_comp} shows the comparison at $z=8$ for BOSS CMASS \citep{Anderson:2012}, which detected the BAO peak at the $5\sigma$ level,
 and \wfirst\ utilizing the number density and bias measurements 
from \bluetides. Although the dark matter power spectrum is an order of magnitude lower at $z=8$, we see that \bluetides\ predicts values of $nP_{\rm gg}$ for \wfirst\ at $z=8$ ($nP_{\rm gg}=1.60-4.87$ ) which are comparable to that of the BOSS CMASS sample ($nP_{\rm gg} = 2.45$). This is due to the fact that the galaxy bias is larger by an order of magnitude and the expected number density at $z=8$ ($n = 1.2 - 3.6\times10^{-4} h^{3} {\rm Mpc}^{-3}$) is nearly the same as the BOSS CMASS number density ($n = 3\times10^{-4} h^{3} {\rm Mpc}^{-3}$). 
The comoving volume for \wfirst\ is shown in Table
\ref{table:boss_wfirst_comp}
 for the same change in redshift as the BOSS CMASS sample, $\Delta z=0.27$,
 centered around $z=8$ to show that these surveys have similar observation volumes (before selection effects). Since the comoving volume of \wfirst\ is larger ($V=3.89 {\rm Gpc}^3$ for BOSS CMASS and $V=4.58 {\rm Gpc}^3$ for \wfirst), and the values of $nP_{\rm gg}$ are very comparable, this means that if sufficiently accurate redshift information was available for the galaxies,
\wfirst\ would be able to detect the BAO signal at $z=8$. As we have noted 
above, in practice, the WFIRST Grism survey as planned 
will only cover one third of the  area of the HLS, and its spectral
coverage of the Lyman break will not include $z=8-9$ in any case.
Our BAO predictions are therefore mostly illustrative, serving to highlight
that there is potentially a large population of galaxies which could be
used to make this cosmological measurement. 
Further work would be needed to decide whether changes to the WFIRST mission
 instrument parameters would be enough to make the measurement
feasible.
 
\begin{figure}
\includegraphics[width = 0.45\textwidth]{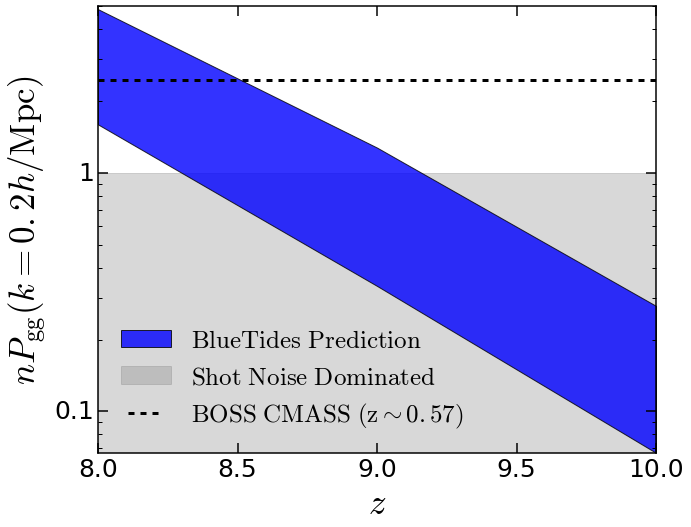}
\caption{$nP_{\rm gg}$ computed for the \wfirst\ survey using the number density and bias computed from \bluetides. Upper (lower) limit uses the number density $n$ computed with the intrinsic (dust corrected) luminosities. Black line shows the value for the BOSS CMASS sample which detected the BAO peak at the $5\sigma$ level. \citep{Anderson:2012}}
\label{fig:nP}
\end{figure}

Figure \ref{fig:nP} shows the product $nP_{\rm gg}$ at the BAO scale
as a function of redshift. The blue band shows $nP_{\rm gg}$ for the
number density of galaxies detectable by \wfirst\ determined by the
intrinsic (upper limit) and dust corrected (lower limit) luminosities.
The black dashed line shows $nP_{\rm gg}$ for the BOSS CMASS sample
\citep{Anderson:2012}. The error on the BAO measurement scales with
$V_{\rm survey}^{-1/2}$ and $(1+nP_{\rm gg})/nP_{\rm gg}$
\citep{Seo:2003pu}. Using this as a rough measure for the detection
strength of the BAO signal compared to the $5\sigma$ detection
strength from the BOSS CMASS sample, \bluetides\ predicts a $\sim
4.7-6.3\sigma$ detection of the BAO signal at $z=8$. At $z=9$ and
$z=10$, \bluetides\ predicts detection significances of $\sim
1.4-4.0\sigma$ and $\sim 0.5-1.5\sigma$, respectively, assuming
accurate redshift information. We find a similar result following the
method of \citet{Blake2003}, including the effects of shot noise, so
that the error on the power spectrum is given by $\sigma_P/P_{\rm gg}
=2\pi( k^2 \Delta kV_{\rm survey})^{-1/2}(1+nP_{\rm gg})/nP_{\rm
  gg}$. Assuming $\Delta k = 0.015 h/Mpc$, we find a $4.7\sigma$
difference between the galaxy power spectrum and the no-wiggle power
spectrum of \citet{Eisenstein1998} at the BAO scale for the non-dust
corrected number density at $z=8$. We note again that inclusion of
uncertainties on photometric redshifts, which will reduce the observed
BAO signal, is necessary for a complete analysis.

\section{Conclusions}
\label{Section: Conclusion}
Using the \bluetides\ cosmological hydrodynamic simulation, we have
forecast the properties of the galaxy and AGN populations to be discovered by
the \wfirst\ HLS in the redshift range $z=8-15$. 
The \bluetides\ simulation produces results which agree well with the galaxy LF from current {\em HST} observations, including the highest redshift galaxy to date \citep{Oesch16}. Our conclusions for the \bluetides\ predictions are as follows:

\begin{itemize}
\item $ \gtrsim 10^6$ star forming galaxies  beyond $z=8$ will be detectable by the \wfirst\ HLS, significantly more than previously predicted (S13).
\item We find that even with a dust-corrected model, the bright end of the luminosity function deviates from that of a standard Schechter function. This is relevant to the \wfirst\ LFs, since the HLS survey will only be able to detect the brightest galaxies at such high redshift. 
\item $z=15$ galaxies are likely to be within the $5\sigma$ detection limit of \wfirst\, since dust effects seem to be small beyond $z=11$.
\item Around $10^4$ AGN will have UV luminosities bright enough to be detected by \wfirst\ at $z=8$ and beyond. These will be the highest redshift AGN 
 observed to date. 
\item The \wfirst\ galaxy population will have specific star formation rates of $\sim 10/ {\rm Gyr}^{-1}$, ages between 10s to 100s of million years, and gas phase metallicities in the range $12+\log_{10}(O/H)=6{\rm\ to\ }8$. The galaxies will reside in dark matter halos with masses $\gtrsim 10^{10.5}M_\odot$. 
\item A few of the brightest AGN sources in the \bluetides\ volume outshine their host galaxy in the UV. The AGN \wfirst\ will observe have black hole
masses that range from $10^5-10^8 M_\odot$ and their mass accretion results in
a luminosity around half the Eddington value or more. 
\item The \wfirst\ galaxy population will be very highly biased. The bias at $z=8$ will be $b=13.4\pm 1.8$ and will evolve linearly with redshift.
\item Due to the high bias and large number density predicted by \bluetides, if redshift information is available for these galaxies 
(using the \wfirst\ Grism)
\wfirst\ could perhaps detect the BAO peak to a high level
of significance at $z=8$, comparable to that of the BOSS CMASS sample. 
We note that this would be unlikely as the mission is currently
planned (requiring changes to the spectral coverage of the Grism). It
is worth considering this result in the future, as it
could lead to high redshift constraints on cosmological model parameters. 
\end{itemize}

The \wfirst\ HLS will result in a huge change in our knowledge of the high redshift observational frontier. The number of observed galaxies will greatly increase: by the most conservative estimates derived from \bluetides, \wfirst\ will increase the $z=8-11$ galaxy sample by a factor of about 10,000 from the current sample and will observe the first galaxies at $z=12$ and beyond. The \wfirst\ observations will further constrain the epoch of reionization, galaxy formation theories, and cosmology in a new redshift regime.

\section{Acknowledgements}
We thank Sebastian Fromenteau for useful discussions.
We acknowledge funding from NSF ACI-1036211,  NSF AST-1517593, NSF AST-1009781, and the BlueWaters PAID program. The
\bluetides\ simulation was run on facilities on BlueWaters at the
National Center for Supercomputing Applications. SMW acknowledges
support from the UK Science and Technology Facilities Council.

\FloatBarrier

\bibliographystyle{mnras}
\bibliography{WFIRST} 

\FloatBarrier

\section{Appendix}
\subsection{DPL Fits to BlueTides LF}

\begin{figure}
\includegraphics[width = 0.45\textwidth]{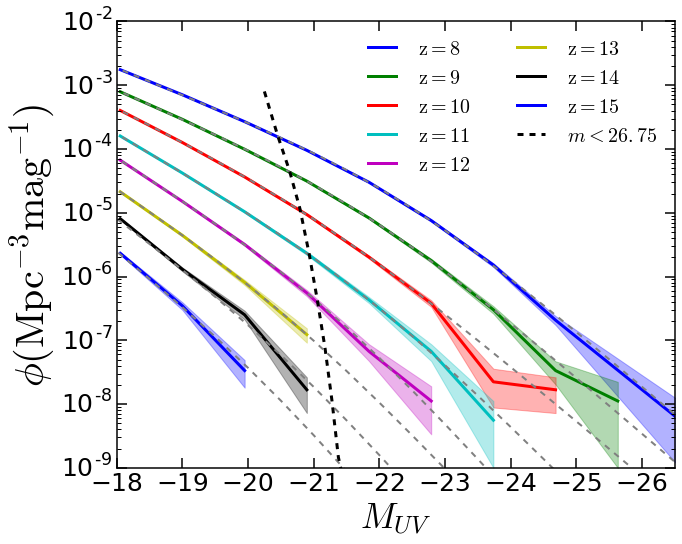}
\caption{Intrinsic luminosity functions for \bluetides. Dashed lines show the best fit DPL defined in equation \ref{dpl}.}
\label{fig:int_lf}
\end{figure}

\begin{figure}
\includegraphics[width = 0.45\textwidth]{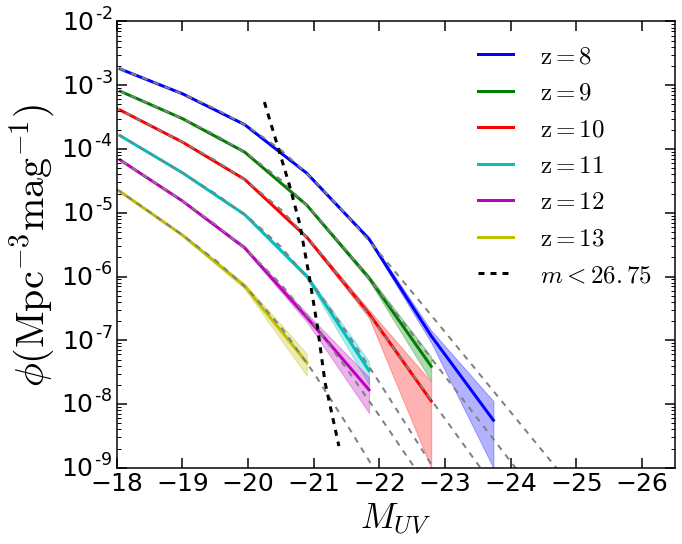}
\caption{Dust corrected luminosity functions for \bluetides. Dashed lines show the best fit DPL defined in equation \ref{dpl}.}
\label{fig:dust_lf}
\end{figure}

In Figure \ref{fig:int_lf} we show the intrinsic luminosity functions for \bluetides\ galaxies with their best fit to the DPL defined in equation \ref{dpl}. Dust corrected luminosity functions are shown in Figure \ref{fig:dust_lf}. We fit the DPL parameters' redshift evolution assuming a linear evolution in $\log(\phi^*), M^*, \alpha, {\rm and\ }\beta$. The results for the intrinsic LFs are:
\begin{eqnarray}
\log(\phi^*) = -\left[(0.96\pm0.22)(z-8)+(2.54\pm2.32)\right]\nonumber \\
M^*= \left[(0.28\pm0.12)(z-8)+(-24.51\pm1.29)\right]\\
\alpha =-\left[(0.14\pm0.02)(z-8)+(0.86\pm0.22)\right] \nonumber\\
\beta =-\left[(0.15\pm0.05)(z-8)+(1.92\pm0.55)\right]  \nonumber
\label{int_fits}
\end{eqnarray}
and for the dust corrected LFs:
\begin{eqnarray}
\log(\phi^*) = -\left[(1.19\pm0.06)(z-8)+(1.13\pm0.62)\right]\nonumber \\
M^*= \left[(0.03\pm0.03)(z-8)+(-20.71\pm0.32)\right]\\
\alpha =-\left[(0.16\pm0.01)(z-8)+(0.66\pm0.15)\right] \nonumber\\
\beta =-\left[(0.22\pm0.06)(z-8)+(2.42\pm0.63)\right]  \nonumber
\label{dust_fits}
\end{eqnarray}
where $\phi^*$ is in $[{\rm Mpc}^{-3} {\rm mag}^{-1}]$. 

We define the AGN power law LF as
\begin{eqnarray}
\log(\phi)=\log(\phi^*)+\log[0.4\log10)] \nonumber
\\+0.4\log(10)(M^*-M)(1-\alpha) 
\end{eqnarray}
where we arbitrarily choose $M^*=-18.0$ (grey dashed lines in Figure \ref{fig:agn_lf}). The best fit redshift evolutions are given by
\begin{eqnarray}
\log(\phi^*) = -\left[(1.44\pm0.26)(z-8)+(26.15\pm2.64)\right]\nonumber \\
\alpha =-\left[(0.01\pm0.07)(z-8)+(2.38\pm0.68)\right] \nonumber\\
\label{agn_fits}
\end{eqnarray}

\subsection{Fits to Cumulative Number Density}

The cumulative number of objects brighter than a given magnitude is well fit by a power law at each redshift (Figure \ref{fig:N_v_m}). We fit the results with the following:
\begin{equation}
\log_{10}\left(N(<m_{\rm UV}) /{\rm deg}^2\right)= A + b\ m_{\rm UV}
\label{N_fit}
\end{equation}
with
\begin{eqnarray}
A = (-2.61 \pm 0.22)(1+z) + (5.93 \pm 2.83)\\
b = (0.07\pm 0.01)(1+z) + (0.19\pm 0.11) \nonumber
\end{eqnarray}
for the intrinsic luminosities and
\begin{eqnarray}
A = (-3.23\pm 0.29)(1+z) + (-1.74\pm 3.67)\\
b = (0.09 \pm 0.01)(1+z) + (0.46 \pm 0.14) \nonumber
\end{eqnarray}
for the dust corrected luminosities.

\FloatBarrier
\label{lastpage}
\end{document}